\documentclass[preprint]{revtex4}
\usepackage{amssymb}
\usepackage{latexsym}
\usepackage{amsmath}
\usepackage{graphicx}
\usepackage{bm}
\usepackage{dcolumn} 

\begin{document}

\title{Summary Talk\\
X Latin American Workshop on Plasma Physics\\
November 30, 2003 - December 5, 2003\\
S\~ao Pedro, S.P., Brazil}
\author{Reuven Opher}
\email{opher@astro.iag.usp.br}
\affiliation{IAG, Universidade de S\~ao Paulo, Rua do Mat\~ ao 1226,\\
Cidade Universit\'aria, CEP 05508-900 S\~ ao Paulo, SP, Brazil}

\begin{abstract}
Of the many important topics that were discussed at the workshop, I summarize and comment on 25 
presentations, which I found to be particularly interesting. They fall into all of the areas
covered in the conference: basic plasma phenomena, space and astrophysical plasmas, technological
applications of plasma, and thermonuclear fusion.
\end{abstract}
\maketitle
\noindent\textbf{I. INTRODUCTION}
\vspace{.7cm}
\par
The Latin American Workshop on Plasma Physics (LAWPP) meets every two years, uniting plasma physicists
from all over Latin America. These workshops also include the participation of physicists from all over the 
world.
\par
There were 40 oral talks and 205 posters presented at the workshop, covering the areas of basic plasma physics 
phenomena, space and astrophysical plasmas, technological applications of plasma, and thermonuclear fusion. 
\par
From among the many excellent presentations, I have chosen to summarize and comment on the following 25, that 
seem to me to be the most interesting, in the limited time available for this summary talk.
\vspace{.8cm}
\par
\noindent\textbf{A. Basic Plasma Phenomena}\\
\vspace{-.6cm}
\begin{enumerate}
\item{Supersonic plumes through magnetized plasmas;}
\item{Coupled map lattices;}
\item{The Kelvin-Helmholtz instability dependence on the structure of the transition layer;}
\item{Stablization of ion-bead instabilities;}
\item{The nature of ball lightning: a force-free magnetic field;}
\item{The nature of ball lightning: coupled vortex-magnetic field;}
\item{Observing plasma temperatures;}
\item{Multi-harmonic Langmuir wave generation in beam-plasma interactions.}
\end{enumerate}
\vspace{.5cm}
\textbf{B. Space and Astrophysical Plasmas}\\
\vspace{-.6cm}
\begin{enumerate}
\item{The effective charge of a neutrino in a plasma;}
\item{The influence of solar plasma density perturbations on neutrino propagation;}
\item{Lightening bead instabilites;}
\item{Gyrophase electron organization in space plasmas.}
\end{enumerate}
\vspace{.5cm}
\textbf{C. Technological Applications of Plasma}\\
\vspace{-.6cm}
\begin{enumerate}
\item{Plasma processing of domestic and industrial waste;}
\item{Space plasma propulsion using the Hall effect.}
\end{enumerate}
\vspace{.5cm}
\textbf{D. Thermonuclear Fusion}\\
\vspace{-.6cm}
\begin{enumerate}
\item{Alfven wave heating in tokamaks;}
\item{Instability supression in a Z-pinch by an axial sheared flow;}
\item{Spark ignition in an inertialy confined plasma;}
\item{Present status of ITER;}
\item{Present status of IGNITOR;}
\item{Reflectometry diagnostics for burning plasmas;}
\item{Pellet injection in toroidal plasmas;}
\item{Controlling turbulence;}
\item{Effects of electric fields on MHD activity;}
\item{Magnetic islands, radial electric fields, and plasma rotation in tokamaks;}
\item{Diffusion of heavy charged ions, radial electric fields, and plasma rotation in tokamaks.}
\end{enumerate}
This summary is organized around the main ideas in the talks and posters which are reviewed. 
The original titles of the presentations as well as all of the names of the authors can be found at the end of the summary.
\newpage
\noindent\textbf{II. SUMMARY AND COMMENTS}\\
\vspace{.001cm}
\par
\noindent\textbf{A. Basic Plasma Phenomena}\\
\vspace{-.2cm}
\par
\noindent\textbf{1. Supersonic plumes moving through magnetized plasmas}\\
Morales and Tsung discussed supersonic plumes moving through magnetized plasmas. 
Supersonic plumes created by laser ablation are generated in the Basic Plasma Facility at UCLA. Investigating the properties of these plumes with an electromagnetic particle-in-cell code, it was found that there is a ballistic expansion of energetic electrons along the field. The return currents excite various electromagnetic waves: shear Alfven waves along the field and Whistler and compressional modes in the transverse direction.\\ 
\vspace{-.45cm}
\par
\noindent{\textbf{Comment on the presentation:}} What is the energy loss rate of the expanding plume as a function of the density and magnetic field of the background plasma?\\
\vspace{-.2cm}
\par
\noindent\textbf{2. Coupled map lattices}\\
Coupled map lattices (CMLs) are spatially extended systems with discrete time and space intervals and a continuous, connected dynamical variable. Vasconcelos and Viana  computed a correlation integral and evaluated a spatial correlation dimension for the lattices.\\
\vspace{-.45cm}
\par
\noindent{\textbf{Comment on the presentation:}} Is it possible to use CMLs to evaluate the correlation of galaxies in astrophysics?\\
\vspace{-.2cm}
\par
\noindent\textbf{3. The Kelvin-Helmholtz (KH) instability dependence on the structure of the transition layer}\\
It was shown by Gratton et al. that, whereas a smooth transition between two uniform regions of 
different densities, velocities, and magnetic fields may be unstable, a discontinuous transition may, in fact, be stable. This is due to the stabilizing effect of magnetic shear. (A discontinuous transition has a very large shear.) A stability diagram of $d/\lambda$ vs $M_{\textrm A}$, where $d$ is the gradient length scale, $\lambda$ is the wavelength of the instability and $M_{\textrm A}$ is the Alfvenic Mach number, was presented.  For a given 
$d$ and $M_{\textrm A}$ as well as a sufficiently large $\lambda$ (i.e., small $d/\lambda$), the region is KH stable.\\
\vspace{-.45cm}
\par
\noindent{\textbf{Comment on the presentation:}} Could this be relevant to astrophysical jets, which are subject to 
KH instabilities?\\
\vspace{-.2cm}
\par
\noindent{\textbf{4. Stablization of ion-beam instabilities}}\\
The ion-beam stream instability is well-known in plasma physics. It was shown by Gomberoff, Hoyos and Brinca
that right-handed electromagnetic instabilities can be stabilized by nonlinear left-handed polarized
waves. Stabilization occurs above a certain threshold, which increases with increasing beam velocity.\\
\vspace{-.45cm}
\par
\noindent{\textbf{Comment on the presentation:}} Can ion-beam plasma instabilities be stabilized by detecting the 
instability that is developing and then applying a sufficiently intense electromagnetic wave of the opposite 
polarization?\\ 
\vspace{-.2cm}
\par
\noindent\textbf{5. The nature of ball lightning: a force-free magnetic field}\\
The phenomenon of ball lightning, a bright glowing ball of hot plasma floating around in a room, has been
reported by many independent observers. It was frequently observed in the battery area of American submarines 
during World War II. Afterwards, it was reproduced in the Los Alamos Laboratory. Tsui suggested that 
the ball is a force-free plasmoid, with the current parallel to the magnetic field in a minimum energy state,
under the constraint of magnetic helicity conservation. The magnetic field decays adiabatically in a sequence
of force-free configurations with lifetimes of many seconds.\\
\vspace{-.45cm}
\par
\noindent{\textbf{Comment on the presentation:}} A high magnetic field is required to produced ball lightning. Can this field be detected experimentally?\\
\vspace{-.2cm}
\par
\noindent\textbf{6. The nature of ball lightning: coupled vortex-magnetic field}\\
Taveira, Sakanaka and Scussiatto studied a coupled vortex-magnetic field solution for ball lightning, 
investigating the triple Beltrami equation. This equation is the sum of four terms with arbitrary constant
coeficients: 1) the magnetic field {\textrm{\textbf{B}}}; 2) the curl of {\textrm{\textbf{B}}}; 3) the curl of the curl of {\textrm{\textbf{B}}}; and 4) the curl of the curl of the curl of {\textrm{\textbf{B}}}. Their solutions to the triple Beltrami equation are able to explain ball lightning.\\
\vspace{-.45cm}
\par
\noindent{\textbf{Comment on the presentation:}} It is assumed that the ball lightning is rotating. Can the rotation be observed experimentally?\\
\vspace{-.2cm}
\par
\noindent\textbf{7. Observing plasma temperatures}\\
Doppler or Stark broadening of spectral lines are generally used to determine the temperature of plasmas.
However, this method is difficult to apply to high temperature plasmas with highly ionized atoms, when 
the spectral lines are in the ultraviolet. Borges et al. studied the relative intensity of spectral
lines as a function of temperature, using a multiple level approach. They calculated the transition 
probabilities in terms of the oscillator strength parameters.\\
\vspace{-.45cm}
\par
\noindent{\textbf{Comment on the presentation:}} It is interesting to compare the methods of Borges et al. for determining the temperature of plasmas with the Doppler and Stark line broadening methods. 
Do both methods give the same temperatures for laboratory plasmas?\\
\vspace{-.2cm}
\par
\noindent\textbf{8. Multi-harmonic Langmuir wave generation in beam-plasma interactions}\\
Gaelzer et al.  discussed the multi-harmonic generation of Langmuir waves from electron beams in plasmas. These
waves are observed in solar flares as well as in the earth's foreshock. A power law dependence of the intensity
on frequency proportional to $\omega^{-\alpha},$ where $\alpha$ varies from 2 to 5, is observed. Five 
harmonics which appear early in the linear growth phase have been observed. A theory was presented to explain 
these observations.\\
\vspace{-.45cm}
\par
\noindent{\textbf{Comment on the presentation:}} Can the theory be tested in the laboratory?\\
\vspace{.01cm}
\par
\noindent\textbf{B. Space and Astrophysical Plasma}\\
\vspace{-.3cm}
\par
\noindent\textbf{1. The effective charge of a neutrino in a plasma}\\
It was argued by Serbeto et al.  that collective plasma effects induce an effective charge on 
neutrinos in dense magnetic plasmas. The effectively charged neutrinos easily couple to plasma 
oscillations. This effect could reactivate the stalled shock in type II supernovae explosions. It can also 
generate electric currents and intense magnetic fields in supernovae cores as well as at the surfaces of 
neutron stars.\\
\vspace{-.45cm}
\par
\noindent{\textbf{Comment on the presentation:}} Can this effect be observed in the interaction of solar neutrinos emitted from the center of the sun with the turbulent plasma near the sun's surface?\\
\vspace{-.2cm}
\par
\noindent\textbf{2. The influence of solar plasma density perturbations on neutrino propagation}\\
There exists a resonance between the g-modes and Alfven waves inside the sun. This resonance can create 
density fluctuations $\delta\rho/\rho$ on the order of 4-8\%. It is known that different types of neutrinos
(electron, muon, $\tau$ meson) mix with one another. The mixing is characterized by the mass difference
and the mixing angle. Larger $\delta\rho/\rho$ indicates a smaller mixing angle. It was shown by Reggiani, 
Guzzo, and Holanda that $\delta\rho/\rho\sim 4-8\%$ can cause observable changes in the mixing angle.\\ 
\vspace{-.45cm}
\par
\noindent{\textbf{Comment on the presentation:}} Can the mixing angle of neutrinos from reactors (i.e., without $\delta\rho/\rho$) be measured sufficiently accurately so that the $\delta\rho/\rho$ in the sun can be determined?\\
\vspace{-.2cm}
\par
\noindent\textbf{3. Lightning bead instabilities}\\
Ludwig studied the origin of the observed beaded structure of decaying lightning strokes.  The 
self-induced magnetic field is unstable to the kink and sausage MHD instabilities. These instabilities lie in
the correct range to explain the spatial structure of the beads.\\
\vspace{-.45cm}
\par
\noindent{\textbf{Comment on the presentation:}} Can the characteristics of the beaded structure of lightning be reproduced in the laboratory in order to test the theory?\\
\vspace{-.2cm}
\par
\noindent\textbf{4. Gyrophase electron organization in space plasmas}\\
The velocity distribution of electrons in the plane perpendicular to the ambient magnetic field is not 
angular independent. This electron nongyrotropy has been suggested as the source of ion acoustic waves in the
solar wind. Moraes and Alves presented solutions of the nongyrophase  dispersion equations.\\
\vspace{-.45cm}
\par
\noindent{\textbf{Comment on the presentation:}} In addition to ion acoustic waves in the solar wind, what are the other possible effects of the electron nongyrotropy?\\
\vspace{.01cm}
\par
\noindent\textbf{C. Technological Applications of Plasma}\\
\vspace{-.3cm}
\par
\noindent\textbf{1. Plasma processing of domestic and industrial waste}\\
Leal-Quiro discussed plasma torch processing of domestic and industrial solid waste. There exist
seven plants in the world, two of which are already in operation, which process waste in this way. The organic 
waste is converted into Synthesis
gas (Syn gas), which is essentially a mixture of $H_2$ and CO. The inorganic components are converted into construction bricks and architectural tiles. The plasma torch operates at a temperature of $\sim 5000^{\circ}C - 100,000^{\circ}C,$ whereas the simple burning of garbage occurs at a temperature $\simeq 2000^{\circ}C.$ At the plasma torch temperatures, the garbage is completely transformed into the tiles and Syn gas, while the simple burning of paper, for example, leaves a residue to 50\% by weight in ash. It has been shown that the initial investment in the processing plant pays for itself within seven years.\\
\vspace{-.45cm}
\par
\noindent{\textbf{Comment on the presentation:}} Why isn't this process used throughout the world?\\
\vspace{-.2cm}
\par
\noindent\textbf{Space plasma propulsion using the Hall effect}\\
Ferreira et al. discussed the Hall Thruster for space propulsion. This propulsion method
has the great advantage of low power consumption when using permanent magnets.\\
\vspace{-.45cm}
\par
\noindent{\textbf{Comment on the presentation:}} Is the Hall Thruster likely to be used for future deep space, long duration satellite missions?\\
\vspace{.01cm}
\par
\noindent\textbf{D. Thermonuclear Fusion}\\
\vspace{-.3cm}
\par
\noindent\textbf{1. Alfven wave heating in tokamaks}\\
Elfimov discussed Alfven wave heating in tokamaks. This process is based on the mode conversion of
compressional waves, created by an external antenna, into shear Alfven waves which heat the plasma. The
conversion takes place at the Alfven resonance layer, where the local Alfven velocity matches the phase velocity 
of the compressional wave. Using an electric cyclotron emission radiometer, local Alfven heating of the plasma in
the Brasilian TCABR tokamak has been observed.\\
\vspace{-.45cm}
\par
\noindent{\textbf{Comment on the presentation:}} Can Alfven wave heating decrease instabilities in tokamaks?\\ 
\vspace{-.2cm}
\par
\noindent\textbf{2. Instability suppression in a Z-pinch by an axial sheared flow}\\
Herreira and Yepez analyzed the studies made on instability suppression in a Z-pinch by shear flow.
The stabilizing effect of shear flow in a Z-pinch has been experimentally observed.\\
\vspace{-.45cm}
\par
\noindent{\textbf{Comment on the presentation:}} Is shear flow the principal stabilizing mechanism in extremely long collimated astrophysical jets, which have been observed?\\
\vspace{-.2cm}
\par
\noindent\textbf{3. Spark ignition in an inertialy confined plasma}\\
It was argued by Bilboa et al. that spark ignition seems to be more promising than volume ignition, based on
their numerical simulations of spark and volume ignition in a D-T plasma, compressed by an imploding liner. 
The objective of their investigation was to amplify the fusion energy in a small
conical channel to a level, at which, detonation can be triggered in a large plasma volume.\\ 
\vspace{-.45cm}
\par
\noindent{\textbf{Comment on the presentation:}} This method is essentially the same as that of a hydrogen bomb. Can the energy be confined in the small spark region so that ignition occurs before it leaks into the entire volume?\\
\vspace{-.2cm}
\par
\noindent\textbf{4. Present status of ITER}\\
Varandas and Novrati discussed the present status of the ITER program. ITER will study $Q\sim 1-10$
(JET has a $Q\sim 0.6),$ where $Q$ is the ratio of the power produced by fusion to the power needed to heat the
plasma. For $Q\sim 10,$ ITER will have a particle density $n\sim 2\times 10^{20}\;\textrm{m}^{-3},$ a temperature 
$T\sim 10\;\textrm{keV},$ a magnetic field $B\sim 5$ tessler and a confinement time $t\sim 1.5$ seconds.\\
\vspace{-.45cm}
\par
\noindent{\textbf{Comment on the presentation:}} Will ITER study all the physics necessary to make a fusion reactor?\\
\vspace{-.2cm}
\par
\noindent\textbf{5. Present status of IGNITOR}\\
Bombarda discussed IGNITOR, which will be able to ignite a plasma. It 
will have a toroidal magnetic field of 13 tessler, a poloidal field of 3.5 tessler and
a confinement time of 0.62 seconds. This is a relatively inexpensive machine, with which to study ignition physics.\\
\vspace{-.45cm}
\par
\noindent{\textbf{Comment on the presentation:}} Although IGNITOR is relatively inexpensive, compared to ITER, as well as being an interesting experiment, it does not replace ITER. While ITER will be used to study fusion reactor physics, the 
purpose of IGNITOR is primarily to achieve ignition. Will it be possible to study the details of the ignition
process with IGNITOR?\\
\vspace{-.2cm}
\par
\noindent\textbf{6. Reflectometry diagnostics for burning plasmas}\\
The diagnostic experiments for ITER present a great challenge since the equipment will have to operate in a harsh
environment for a long time. Manso reviewed the present status of reflectometry, which is a robust diagnostic,
capable of making a wide range of plasma measurements. In particular, it can be used for density 
distribution studies as well as for the study of edge-localized modes (ELMs).\\
\vspace{-.45cm}
\par
\noindent{\textbf{Comment on the presentation:}} Can the ELMs detected by reflectometry be controlled by auxiliary 
heating?\\
\vspace{-.2cm}
\par
\noindent\textbf{7. Pellet injection in toroidal plasmas}\\
The complete interaction between a pellet and a plasma has not been throughly studied until now. Sato
discussed our present knowledge about pellet injection. Fast density oscillations have been
observed just after pellet injection. A long helical tail, which is independent of the magnetic field, 
has been observed.\\
\vspace{-.45cm}
\par
\noindent{\textbf{Comment on the presentation:}} Can pellet injection suppress instabilities and/or create transport barriers?\\
\vspace{-.2cm}
\par
\noindent\textbf{8. Controlling turbulence}\\
In general, plasma edge turbulence should be avoided since it is responsible for increasing particle
transport, driving particles out of the plasma and decreasing confinement. Baptista et al. 
studied the effect of Alfven waves on the structures present in the plasma edge turbulence. They 
showed that the injection of Alfven waves in the Brazilian tokamak, TCABR, decreases the plasma edge turbulence,
though not completely destroying it.\\ 
\vspace{-.45cm}
\par
\noindent{\textbf{Comment on the presentation:}} Is it possible to create Alfven waves that completely destroy the plasma edge turbulence?\\
\vspace{-.2cm}
\par
\noindent\textbf{9. Effects of electric fields on MHD activity}\\
It is well known that electric fields in a tokamak can create H-modes, which are important in plasma 
confinement. The effects of electric fields on MHD activity, however, have not been appreciably studied.
In the previous Brazilian tokamak, TBR-1, Nascimento et al. found that an applied radial electric field
increased the periods of rotation of the m=1, 2, and 3 magnetic islands by a factor of 2.\\ 
\vspace{-.45cm}
\par
\noindent{\textbf{Comment on the presentation:}} Can we control the instabilities and transport in tokamaks by applying radial electric fields?\\
\vspace{-.2cm}
\par
\noindent\textbf{10. Magnetic islands, radial electric fields, and plasma rotation in tokamaks}\\
In the nonlinear stage of the tearing mode instability, magnetic islands are created in toroidally
confined plasmas. Magnetic islands in tokamaks are characterized by their width and rotation frequency.
Plasma rotation is connected with the radial electric field in the plasma column. Sheared radial electric
fields can create transport barriers. Severo et al. studied these effects in the Brazilian
tokamak, TCABR. They found that the plasma rotates with the magnetic islands and that the 
rotation velocity agrees with neoclassical theoretical predictions, within the error limits.\\
\vspace{-.45cm}
\par
\noindent{\textbf{Comment on the presentation:}} To what accuracy can the neoclassical theory predict the width of the magnetic islands and their rotation velocities?\\
\vspace{-.2cm}
\par
\noindent\textbf{11. Diffusion of heavy charged ions, radial electric fields, and plasma rotation in tokamaks}\\
Using a Hasekawa-Wakatani (HW) potential for a turbulent plasma, Tendler predicted a strong diffusion
of $C^+$ ions. The charge separation creates a radial electric field on the order of kV/m. Plasma
rotation results due to the $E\times B$ drift.\\
\vspace{-.45cm}
\par
\noindent{\textbf{Comment on the presentation:}} Can this theory be tested quantitatively?\\
\vspace{.2cm}
\par
\noindent\textbf{III. AUTHORS AND TITLES OF PRESENTATIONS}\\
\vspace{.001cm}
\par
\noindent{M. S. Baptista, I. L. Caldas, M. V. A. P. Heller, and A. A. Ferreira\\}
``Periodic Driving of Plasma Turbulence".\\
\par\noindent
L. Bilbao, G. Linhart, and L. Bernal\\
``Ignition in an Inertially Confined Z-Pinch".\\
\par\noindent
F. Bombarda\\
``The IGNITOR Project".\\
\par\noindent
F. O. Borges, G. H. Cavalcanti, N. L. P. Mansur, and A. G. Trigueiros\\  
``Determination of Plasma Temperature by a Semi-Empirical Method".\\
\par\noindent
A. Elfimov\\
``Results of Alfven Wave Heating in TCABR".\\
\par\noindent
J. L. Ferreira, D. A. Raslan, G. M. de. Carvalho, J. Cezario, M. Junior, R. B. Santiago, I. do 
Rego, and I. S. Ferreira\\ 
``Plasma Diagnostics and Performance of Permanent Magnet Hall Thruster".\\  
\par\noindent  
R. Gaelzer, P. H. Yoon, T. Umeda, and Y. Omura\\
``Harmonic Langmuir Wave Generation in Weak Beam-Plasma Interaction".\\
\par\noindent
L. Gomberoff, J. Hoyos, and A. Brinca\\
``The Effect of a Large Amplitude Circularly Polarized Wave on Linear 
Beam-plasma\\ Electromagnetic Instabilities".\\
\par\noindent
F. T. Gratton, G. Gnavi, L. Bender, and C. Farrugia\\ 
``On the MHD Boundary of the Kelvin-Helmholtz Stability Diagram at Large Wavelengths".\\
\par\noindent
J. J. E. Herrera and M. Y. Y\'epez\\
``Instability Supression by Axial Sheared Flow in Dense Z-Pinch Devices".\\
\par\noindent
E. Leal-Quiros\\ 
``Plasma Processing of Municipal Solid Waste".\\
\par\noindent
G. O. Ludwig\\
``Macroscopic Instabilities in Lightning".\\
\par\noindent
M. E. C. Manso\\
``Reflectometry Diagnostics for Burning Plasma Experiments".\\
\par\noindent
G. J. Morales and F. S. Tsung\\
``Dynamics of Supersonic Plumes Moving Through Magnetized Plasmas".\\
\par\noindent
I. C. Nascimento, I. E. Chamaa Neto, Y. Kouznetsov, V. S. Tsypin, and J.H.F. Severo\\
``Influence of Electric Fields on Plasma MHD Activity in a Small Tokamak".\\
\par\noindent
G. Navrati\\
``Assessment made of the FiRE, IGNITOR, and ITER projects at the Snowmass Meeting".\\
\par\noindent
N. Reggiani, M. M. Guzzo, P. C. de Holanda\\
``Influence of the Solar Density Perturbations on the Neutrino Propagation".\\
\par\noindent
K. Sato\\
``Studies on Various Phenomena During the Ice Pellet Injection into Toroidal Plasmas".\\
\par\noindent
A. de P. B. Serbeto, L. A. Rios, P. K. Shukla, and J. T. Mendon\c ca\\
``Neutrino Effectivo Charge in a Magnetized Plasma".\\
\par\noindent
J. H. F. Severo, J. C. Nascimento, V. S. Tsypin, R. M. O. Galv\~ao, Y. K. Kuznetsov, E. A.\\ 
Saettone, A. Vannucci, A. B. Mikhailovskii, and M. Tendler\\
``Magnetic Islands and Plasma Rotation in TCABR Tokamak".\\
\par\noindent
A. M. A. Taveira, P. H. Sakanaka, and C. E. Scussiatto\\
``Coupled Vortex-Magnetic Field Solution for Ball Lightning".\\
\par\noindent
M. Tendler\\
``Physics of the Transport Barriers".\\
\par\noindent
K. H. Tsui\\
``Ball Lightning as a Magnetostatic Spherical Force-Free Field Plasmoid".\\ 
\par\noindent
C. Varandas\\
``Overview of the European Fusion Program with Emphasis on the ITER Activity,\\ 
Leading to the Study of Burning Plasmas".\\
\par\noindent
D. B. Vasconcelos and R. L. Viana\\
``Recurrence Plots in the Investigation of Spatial Behaviour in Coupled Map Lattices".\\
\vspace{-1cm}
\par
\acknowledgements
The author would like to thank the organizing committee for inviting him to give this summary talk and
the Brazilian financing agencies FAPESP (00/06770-2) and CNPq (300414/82-0) for partial support.
\end{document}